\documentclass[aps,prl,twocolumn]{revtex4}
\usepackage{epsfig}

\begin{document}

\title{Is a multiple excitation of  a single atom equivalent to a single excitation of an ensemble of atoms?}

\author
{Ido Kanter, Aviad Frydman and Assaf Ater}
\address
{Minerva center and the Department of Physics, Bar Ilan
University, Ramat-Gan 52900, Israel}

\begin{abstract}
Recent technological advances have enabled to isolate, control and
measure the properties of a single atom, leading to the
possibility to perform statistics on the behavior of single
quantum systems. These experiments have enabled to check a
question which was out of reach previously: Is the statistics of a
repeatedly  excitation of an atom N times equivalent to a single
excitation of an ensemble of N atoms? We present a new method to
analyze quantum measurements which leads to the postulation that
the answer is most probably no. We discuss the merits of the
analysis and its conclusion.

\end{abstract}
\maketitle

The development of laser cooling techniques have enabled to study
the properties of single atoms. This research is motivated both by
the quest for better understanding of basic quantum mechanics
concepts as well as by potential applications in the fields of
quantum computers, quantum clocks and random number generators.
\cite{wieman,berkeland,itano}. Though the accuracy of quantum
mechanical experiments is increasing rapidly, the analysis of the
data is still in its primary stage and a method for detecting
correlations is far from being established.

An example for the above effort is the advanced study of quantum jumps
in a single atom. Figure 1 shows the results of a typical quantum jump
experiment performed on a $^{189}Hg^{+}$ ion (data provided by W.M Itano).
The atomic system contains two excited states (as seen in figure 2).
State 1 has a strong coupling to the ground state while state 2 is
weakly coupled and has a long lifetime. If the system is excited to
state 1 it will emit a photon as the system decays to the ground
state (``light'' level) unless it is transferred to state 2 in which
case the fluorescence will stop for a period of the lifetime of this
state (``dark'' level). The data is viewed as a set of switches between
two levels. The higher one representing detection of emitted photons as
the sample decays from state 1 to the ground state and the lower one
(basically 0) represents the situation where the sample is at state 2.

\vspace{1.5cm}
\begin{figure}
\centerline{\epsfxsize=3.25in \epsffile{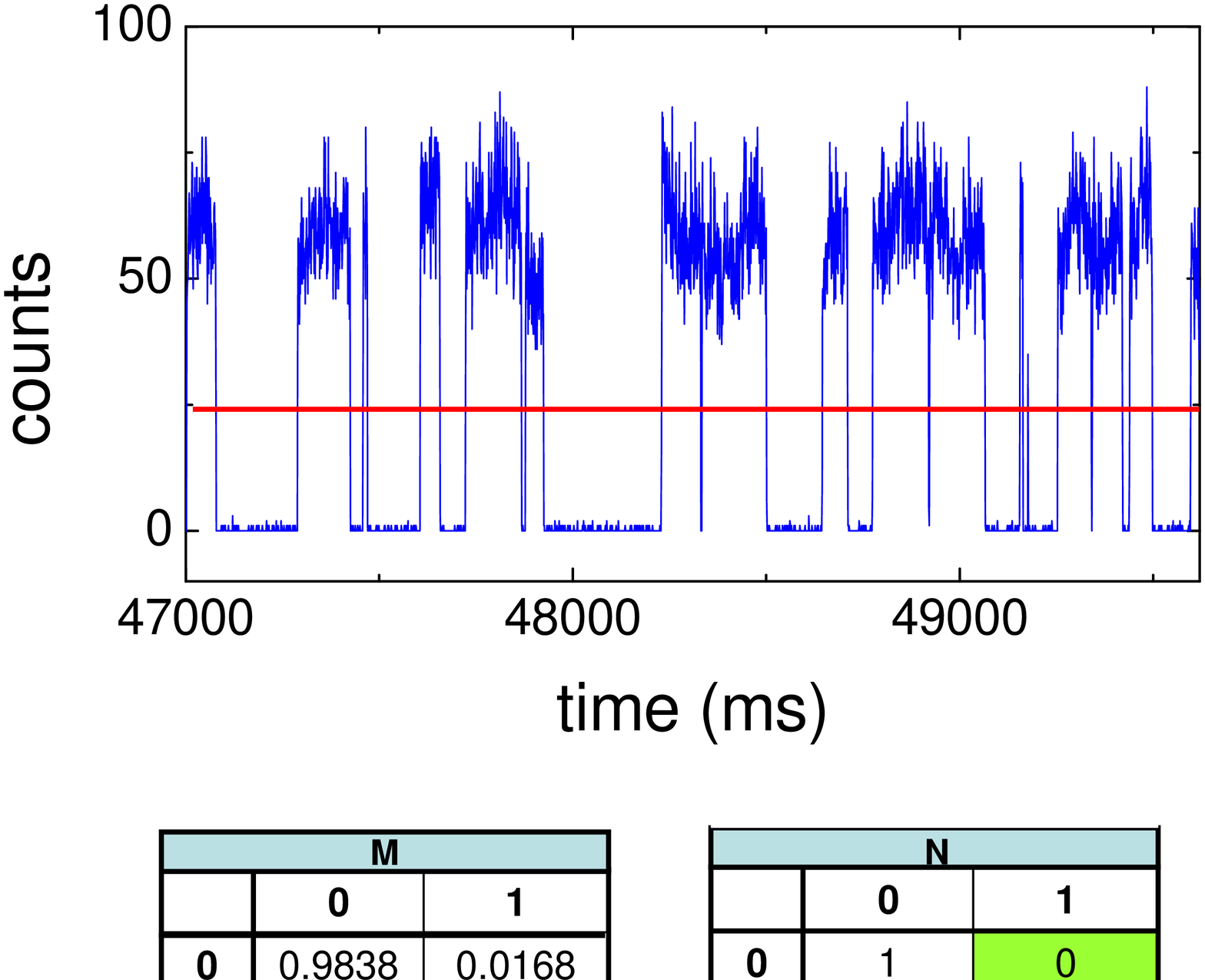}} \vspace{1.5cm}
\caption{A section of a typical quantum jump measurement performed
on a Mercury ion. The red line shows the clipping threshold for
the analysis process (results are insensitive to the precise
threshold value). The figure also depicts the Markov matrix (M)
and the noise matrix (N) obtained from the analysis of the entire
experimental data set using by the BW algorithm. } \end{figure}

This type of experiments opens the opportunity to examine new
questions related to quantum mechanics. It is well established
that the decay of an ensemble of quantum particles is governed by
a Poissonian process. The number of decay events as a function of
time is proportional to $exp(-t/\tau)$, where $\tau$ is the
relevant life time. This behavior indicates that the process is
random with no correlations among successive events. Such
statistics were confirmed in many quantum experiments such as
Alpha and Beta radiation and decay of excited atoms to the ground
state \cite{ensemble}. The question we raise in this work is
whether the excitation of a single atom many times would exhibit
the same behavior and statistics. For instance, in the quantum
jumps experiments, the question is whether the level lifetimes are
random and temporally uncorrelated. This question is of major
importance both for the foundation of quantum mechanics as well as
for setting guidelines for future applications.

\vspace{0.0cm}
\begin{figure}
\centerline{\epsfxsize=3.25in \epsffile{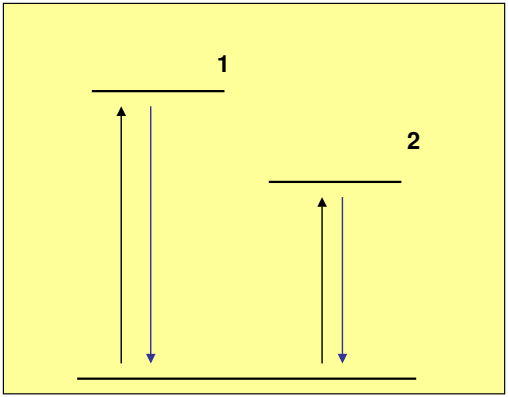}} \vspace{1.0cm}
\caption{A schematic energy diagram of the ionic system. }
\end{figure}

In order to check this question we develop a method to analyze the
data of quantum jumps experiments. This method is based on the
theory of the hidden Markov model\cite{BW,2,paper}. A Markov model
is a finite state machine that changes state once every time unit.
The manner in which the state transitions occur is probabilistic
and is governed by a state-transition matrix, M. For applying this
model to quantum jumps we focus on the simplest case of a Markov
model, i.e. only two levels ``light'' and ``dark''.  Quantum jumps
are expected to be described by a Markovian process, since events
are assumed to depend only on the current state and not on the
history.  If on the other hand, a system does not follow pure
Markovian statistics, the process is named a Hidden Markov
Process. For example, if the atom experiences external fields or
interactions, one may expect to observe a deviation from Markovian
behavior. In this case the process has to be described by two
$2X2$ transition matrices. The first one, M, stands for the
transitions due to the Markovian process (M(i,j) stands for the
transition from i to j) while the second one, N, represents
unexpected transitions that can not be ascribed to the usual
statistics of the system.  The off-diagonal elements of matrix N,
known also as the noise, measure the amount of deviation from pure
Markovian behavior. For pure Markovian processes the off-diagonal
elements are zero and as these elements increase the noise
increases. A practical way to determine the two matrices, M and N,
that characterize the most likely underlying processes, is known
as the Baum-Welch (BW) algorithm \cite{BW}.

The actual analysis of the quantum jumps data is performed using
the following procedure.  A data set, such as that of figure 1, is
clipped to produce a symbolic data sequence containing two levels,
0 and 1 (0 representing a "dark" level and 1 representing the
``light'' level). The transitions between these levels are
analyzed by running the Baum-Welch procedure \cite{BW} to produce
two matrices, M, and, N. A typical result is shown in figure 1.
Here we analyze the data of a sequence of $10^5$ data points where
each point represents the number of photon counts in a time
interval of $1ms$.

The off-diagonal elements of matrix N show that while there are no
unexpected transitions from $0 \rightarrow 1$, there are
non-trivial transitions from $1 \rightarrow 0$ that are not
consistent with a simple Markovian framework. It is important to
note that though the effect is rather small, it can not be
attributed to a finite size artifact. We have generated artificial
sequences (using matrix M only) having sizes similar to the
experimental one which produced pure Markov matrices with
negligible off-diagonal elements of matrix N. Similar
non-Markovian results were obtained on a longer sequence
($7.5*10^5$ data points) taken from a Sr quantum jumps experiment
\cite{berkeland}.

The BW procedure indicates that a single ion produces unexpected
transitions. One may ask whether it is possible to go one step
further and to identify the location (in the time sequence) of the
noise. Such a question arises, for instance, in the case of
digital communication, where a Markovian message is transmitted
through a noisy channel. The receiver's goal is to identify the
locations of the noisy bits in order to recover the original
signal \cite{cover}. An established way to do this is by using the
Viterbi algorithm \cite{viterbi1} to compute the most likely
underlying Markov sequence, or, in other words, the expected
physical outcome without the detected noise \cite{viterbi}.

The outcome of running the Viterbi algorithm on the data of figure 1 is presented in figure 3.  The clipped data contains sub-sequences like ...000010000... (type A) or ...111101111....  (type B).  The Viterbi procedure identifies most of events of type B as noise, but non of type A. A similar effect was obtained in all studied quantum jump experimental sequences (over 10 data sets).

\vspace{1.5cm}
\begin{figure}
\centerline{\epsfxsize=3.25in \epsffile{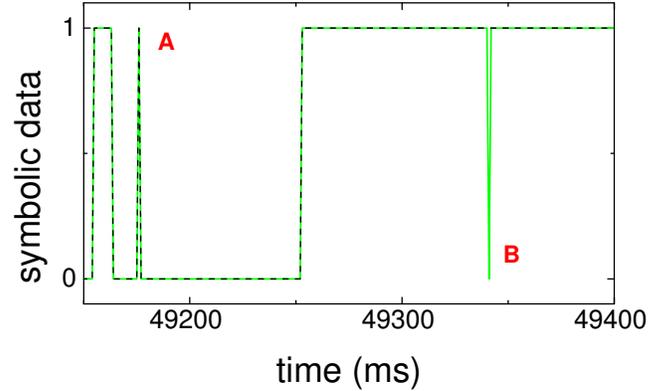}} \vspace{1.0cm}
\caption{Part of the symbolic data sequence after clipping the
data of figure 1. The dashed line shows the result of the Viterbi
algorithm.}
\end{figure}

The above result reveals the fact that sawtooth-like features of type B
(a few dozens in a sequence of $10^{5}$ data points), can not be explained
as part of the usual statistics of the ion, i.e.  Markovian statistics.
We demonstrate that these sawteeth are the cause for non-Markovian noise by
flipping these particular data points from 0 to 1 and running the BW procedure
on the revised sequence. This procedure yields practically pure Markov matrices
even for flipping only $\sim$40\% of the features detected by the Viterbi process as problematic.

It is important to note that deviations from pure exponential behavior of the plateau statistics were observed in quantum jumps experiments \cite{berkeland}. These were attributed to experimental difficulties in controlling the laser amplitude and frequency over relatively large time periods. This experimental artifact gives rise to non-Poissonian statistics due to long plateaus (with times larger than 130 ms).  Our analysis (the BW and the Viterbi algorithms) implies that it is the short plateaus, rather then the long ones, which are responsible for the unexpected behavior.  Further support is obtained by running the BW algorithm on quantum jump data once while erasing long plateaus (having time scales above 100ms) and again while erasing plateaus shorter than 2ms. In the former case we found that the off-diagonal noise remained non-zero. On the other hand the latter case revealed that the noise was entirely suppressed when the short plateau were removed from the sequence. Our analysis therefore strongly implies that the sources of non-Markovian behavior in a single atom experiment are the short plateaus where the system spends a relatively short time at the excited state.

In an attempt to understand the results of the analysis we note
that though an atom is a single quantum particle it is a many body
system containing many degrees of freedom. Therefore, the decay of
the atom to the ground state may have some characteristic
timescales. This might be a source for our observations. If the
ion is repeatedly excited with very short time intervals it may
not be able to fully relax. This may give rise to a deviation from
the expected Markovian statistics. In this respect the excitation
of a single atom many times may differ from a single excitation of
many atoms. A natural way to further check this hypothesis
experimentally is to enhance the effect by taking measurement with
a shorter characteristic timescale (counting photons in time
intervals shorter than a ms). We suspect that such measurements
would result in a larger deviation from Markovian behavior.

We gratefully acknowledge W.M Itano and D. J Berkeland for
providing the quantum jump experimental data and P.W. Anderson and
L. Khaikovich for useful discussions.


\begin{thebibliography}{9}
\bibitem{viterbi} Note that the separation of the received message to signal and noise is not always possible due to either theoretical bounds of information theory or limitations of the used algorithm and the examined sequence. However, in many cases the Viterbi algorithm proves very usuful.
\end{thebibliography}

\end{document}